\documentclass[iop]{emulateapj}
\usepackage{apjfonts,amsmath}
\usepackage[normalem]{ulem}

\shorttitle{Quasar Proper Motions}
\shortauthors{Darling}

\begin{document}
\title{Objects Appear Smaller as They Recede:  How Proper Motions Can Directly Reveal the Cosmic  
Expansion, Provide Geometric Distances, and Measure the Hubble Constant} 

\author{ Jeremy Darling\altaffilmark{1}}
\altaffiltext{1}{Center for Astrophysics and Space Astronomy,
Department of Astrophysical and Planetary Sciences,
University of Colorado, 389 UCB, Boulder, CO 80309-0389, USA; jdarling@colorado.edu}

\begin{abstract}
 Objects and structures gravitationally decoupled from the Hubble expansion will appear to shrink in angular size as the universe
  expands.  Observations of extragalactic proper motions can thus directly reveal the cosmic expansion.
  Relatively static structures such as galaxies or galaxy clusters can potentially be used to measure the Hubble constant,
  and test masses in large scale structures can measure 
  the overdensity.
  Since recession velocities and angular separations can be precisely measured, 
  apparent proper motions can also provide geometric distance measurements to static structures.
  The apparent fractional angular compression 
  of static objects is 15 $\mu$as~yr$^{-1}$ in the local universe; this motion is modulated by the overdensity in
  dynamic expansion-decoupled structures.
  We use the Titov et al.\ quasar proper motion catalog to examine the pairwise proper motion of a sparse 
  network of test masses.  
  Small-separation pairs ($< 200$ Mpc comoving) are too few to measure the expected effect, yielding an 
  inconclusive $8.3\pm14.9$~$\mu$as~yr$^{-1}$.
  Large-separation pairs (200--1500 Mpc) show no net convergence or divergence for $z<1$, 
  $-2.7\pm3.7$~$\mu$as~yr$^{-1}$, consistent with pure Hubble expansion and 
  significantly inconsistent with static structures, as expected.  
  For all pairs a ``null test'' gives $-0.36\pm0.62$~$\mu$as~yr$^{-1}$,
  consistent with Hubble expansion, and excludes a static locus at $\sim$5--10$\,\sigma$ significance 
  for $z\simeq0.5$--2.0.  
  The observed large-separation pairs
  provide a reference frame for small-separation pairs that will significantly deviate
  from the Hubble flow.  
  The current limitation is the number of small-separation objects with precise
  astrometry, but {\it Gaia} will address this and will likely detect the cosmic recession.
\end{abstract}
\keywords{astrometry --- cosmological parameters ---  cosmology: miscellaneous --- cosmology: observations --- 
distance scale --- large-scale structure of universe}

\section{Introduction}

Structures that have decoupled from the Hubble flow will show streaming motions that, while 
straightforward to detect as Doppler shifts along the line of sight, are difficult to distinguish 
from the Hubble expansion itself without an independent distance measure.  
Streaming motions across the line of sight \citep{nusser12}, or simply structures decoupled from the Hubble flow, 
however, are separable from 
the Hubble expansion because no proper motion will occur in a homogeneous expansion.   
Thus, with adequate astrometric precision, one can employ quasars as test masses to both 
detect structures that have decoupled from the Hubble flow (thus measuring masses)
and to directly confirm the homogeneity of the Hubble expansion on large scales.  If one can 
identify high brightness temperature light sources in fairly static structures, such as individual 
galaxies or galaxy clusters, then it is possible to obtain geometric distances and a measurement 
of the Hubble constant from observations of real-time recession.

The apparent size of ``cosmic rulers'' as a function of redshift is a canonical cosmological 
test, but the real-time change in the apparent size of such rulers caused by the cosmic expansion 
has not been explored.  
Here we examine the notion that gravitationally bound objects 
appear smaller as they recede, we develop the method by which this effect can be measured, 
and we apply this technique to extant proper motion data.   
We assume $H_\circ=72$~km~s$^{-1}$~Mpc$^{-1}$
and a flat cosmology with $\Omega_\Lambda = 0.73$ and $\Omega_M = 0.27$.

\begin{figure*}
\epsscale{0.9}
\plotone{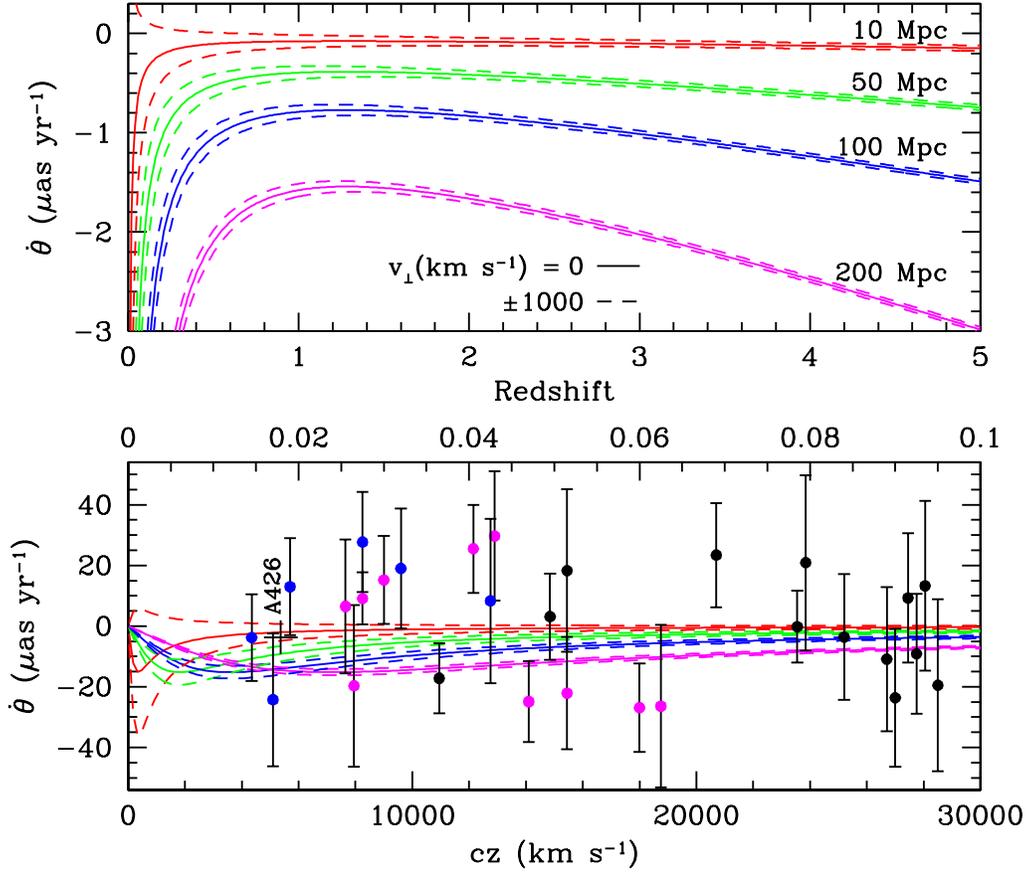}
\caption{\footnotesize
``Observer's plot'' of the expected apparent recession of cosmic structures vs.\ redshift.  
The lower panel shows an expanded view of low redshifts.
Red, green, blue, and magenta lines 
depict structures 10, 50, 100, and 200 Mpc in physical projected size.  
Solid and dashed lines indicate plane-of-sky rest-frame velocities 
$v_\perp = 0$ and $\pm$1000~km~s$^{-1}$, respectively.  
The cross shows the expected effect for the nearby cluster Abell 426, the Perseus Cluster \citep{struble99,hamden10}.
We assume that Abell 426 is completely decoupled from the Hubble
expansion and static.  In reality the apparent $\dot{\theta}$
would be modulated by Equation (\ref{eqn:pm2}).  
Circles show 
measured values of the smallest-separation quasar pairs in the \citet{titov11b} sample with
$|\dot{\theta}|<30$~$\mu$as~yr$^{-1}$ and $\sigma_{\dot{\theta}}<30$~$\mu$as~yr$^{-1}$
(Section \ref{sec:data}), color-coded to match the projected size loci (black points have comoving
separation 250--1500~Mpc).  
All points are consistent with $\dot{\theta} = 0$ 
and none are inconsistent with 
the expected $\dot{\theta}$ for their separation (deviations are less than 3$\,\sigma$).
}\label{fig:thetadot}
\end{figure*}
 
\begin{figure*}
\epsscale{0.9}
\plotone{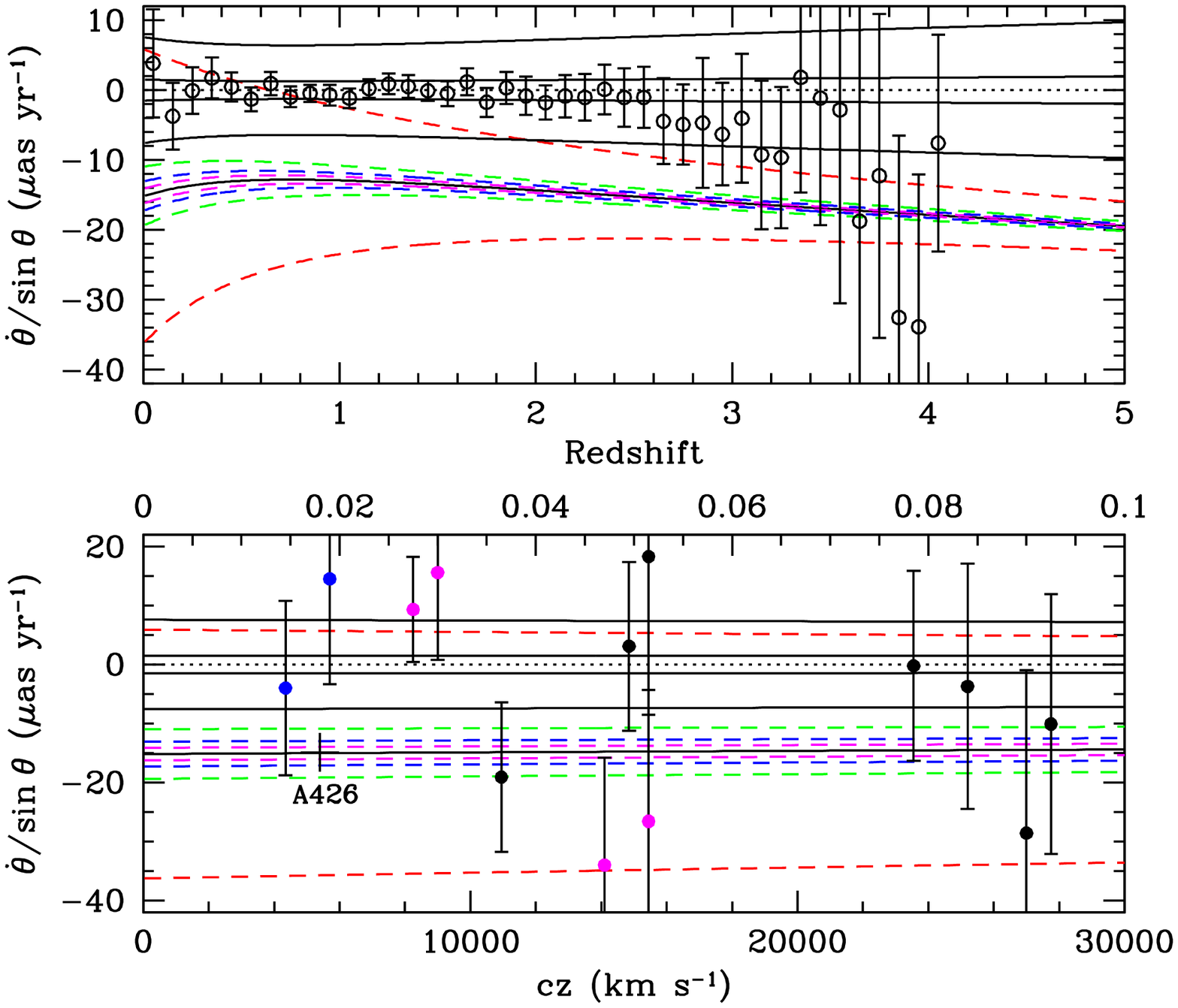}
\caption{\footnotesize
``Theorist's plot'' of fractional apparent recession of cosmic structures vs.\ redshift.  
The lower panel shows an expanded view of low redshifts. Red, green, blue, and magenta lines 
depict plane-of-sky rest-frame velocities $v_\perp = \pm$1000~km~s$^{-1}$ for 
structures 10, 50, 100, and 200 Mpc in physical projected size.  
The dotted black line shows pure Hubble expansion ($\xi=0$, Equation (\ref{eqn:pm2})), 
and solid black lines indicate, from top to bottom, $\xi=-0.5$, $-0.1$, $+0.1$, $+0.5$, and $+1$.
The cross shows the expected effect for the nearby Perseus Cluster, Abell 426 (see the caption to Figure \ref{fig:thetadot}).
Filled circles show 
measured values of the smallest-separation quasar pairs in the \citet{titov11b} sample with
$|\dot{\theta}|<30$~$\mu$as~yr$^{-1}$ and $\sigma_{\dot{\theta}}<30$~$\mu$as~yr$^{-1}$
(Section \ref{sec:data}), color-coded to match the projected size loci (black points have comoving
separation 250--1500~Mpc).  
All points are consistent with $\dot{\theta} = 0$ and none are inconsistent with 
the $\dot{\theta}/\sin\theta$ expected for their separation (deviations are less than 3$\,\sigma$). 
Open circles show a ``null'' test for all pairs, dominated by large-separation pairs ($>1500$ Mpc), which is 
consistent with pure Hubble expansion and excludes static structures, as expected, at 
$\sim$5--10$\,\sigma$ significance for $z\simeq0.5$--2.
}\label{fig:thetadot-theory}
\end{figure*}

\vspace{30pt}

\section{Apparent Proper Motion of Cosmic Rulers}

Given the definition of angular diameter distance, $\theta = \ell/D_A$, where a ``ruler'' of proper length $\ell$ subtends
small angle $\theta$ at angular diameter distance $D_A$, 
both cosmic expansion and a changing $\ell$ can produce an 
observed fractional rate of change in $\theta$:
\begin{equation}
{\Delta\theta/\Delta t_\circ\over\theta} \equiv {\dot{\theta}\over\theta} = { -\dot{D_A}\over D_A} + {\dot{\ell}\over\ell}
= {- H(z)\over1+z} + {\dot{\ell}\over\ell},
\label{eqn:pm}
\end{equation}
where 
\begin{equation}
  H(z) = H_\circ \sqrt{\Omega_{M,\circ}(1+z)^3+\Omega_\Lambda},
\label{eqn:Hofz}
\end{equation}
$\Delta t_\circ$ is the 
observer's time increment, $\dot{\theta}$ is the proper motion, and $\dot{\ell}$ is the observed 
change in proper length, $\Delta\ell/\Delta t_\circ$, related to the physical (rest-frame) 
transverse velocity as $v_\perp=\dot{\ell}\, (1+z)$.
When the small-angle approximation is not valid, we assume that $D_A$ is the angular diameter
distance to the midpoint of $\ell$ such that $\tan(\theta/2) = (\ell/2)/D_A$.  For large angles, Equation (\ref{eqn:pm}) 
becomes
\begin{equation}
{\Delta\theta/\Delta t_\circ\over \sin\theta} = {\dot{\theta}\over\sin\theta} = {- H(z)\over1+z} + {\dot{\ell}\over\ell}\ .
\label{eqn:pm_exact}
\end{equation}  
All calculations use this exact relationship. 

If $\ell$ is not a gravitationally influenced structure and grows with the expansion, then
\begin{equation}
{\dot{\ell}\over\ell} = {H(z)\over1+z}\ , 
\end{equation}
exactly canceling the first term in Equation (\ref{eqn:pm}).  In this case, $\dot{\theta}= 0$, and there
is no proper motion for objects co-moving with an isotropically expanding universe, as expected.  
If $\ell$ is decoupled from the expansion, however, then for most reasonable gravitational motions, 
$\dot{\ell}/\ell$ is a minor modification to the expansion contribution to $\dot{\theta}/\theta$
because the expansion, except for small redshifts or small structures, dominates (Figures \ref{fig:thetadot} and \ref{fig:thetadot-theory}).

This ``receding objects appear to shrink'' observation does not rely on knowledge of the 
orientation or size of the ``ruler'' --- any
relative proper motion between objects that are coupled via gravity will allow a measurement
of $\dot{\theta}$ because the measurement is differential. 

Practically, this effect would be measured via the relative proper motion of high brightness temperature 
light sources such as quasars or masers.  
The convergence (or divergence) of a pair of test masses can be measured via 
\begin{equation}
\dot{\theta}_{12} = -\left(\boldsymbol{\mu}_2\cdot \hat{\theta}_{21} + \boldsymbol{\mu}_1 \cdot \hat{\theta}_{12}\right),
\label{eqn:thetadot_pm}
\end{equation}
where $\boldsymbol{\mu}_i$ is the proper motion, and $\hat{\theta}_{12}$ and $\hat{\theta}_{21}$ are the unit 
vectors connecting the two test masses along a geodesic (the bearing from mass 1 to mass 2 and vice-versa).
The angular separation of two points on a sphere, in terms of right ascension ($\alpha$) and declination ($\delta$), is
\begin{equation}
  \theta_{ij} = \arccos\left(\sin\delta_i\,\sin\delta_j+\cos\delta_i\,\cos\delta_j\,\cos[\alpha_i-\alpha_j]\right),
\label{eqn:theta}
\end{equation}
and this separation changes with time as
\begin{eqnarray}
 	      \dot{\theta}_{ij}= -\left(\cos\delta_i\,\sin\delta_j \left[\mu_{\delta,i}-\mu_{\delta,j}\cos(\alpha_i-\alpha_j)\right]\right. \nonumber\\
                 \left.+\sin\delta_i\,\cos\delta_j \left[\mu_{\delta,j}-\mu_{\delta,i}\cos(\alpha_i-\alpha_j)\right]\right. \nonumber \\ 
                \left. -\cos\delta_i\,\cos\delta_j\,\sin(\alpha_i - \alpha_j)\left[\mu_{\alpha,i}\ - \mu_{\alpha,j}\right]\right)/ \sin\theta_{ij}.
\label{eqn:thetadot_detail}
\end{eqnarray}

\begin{figure*}
\epsscale{0.9}
\plotone{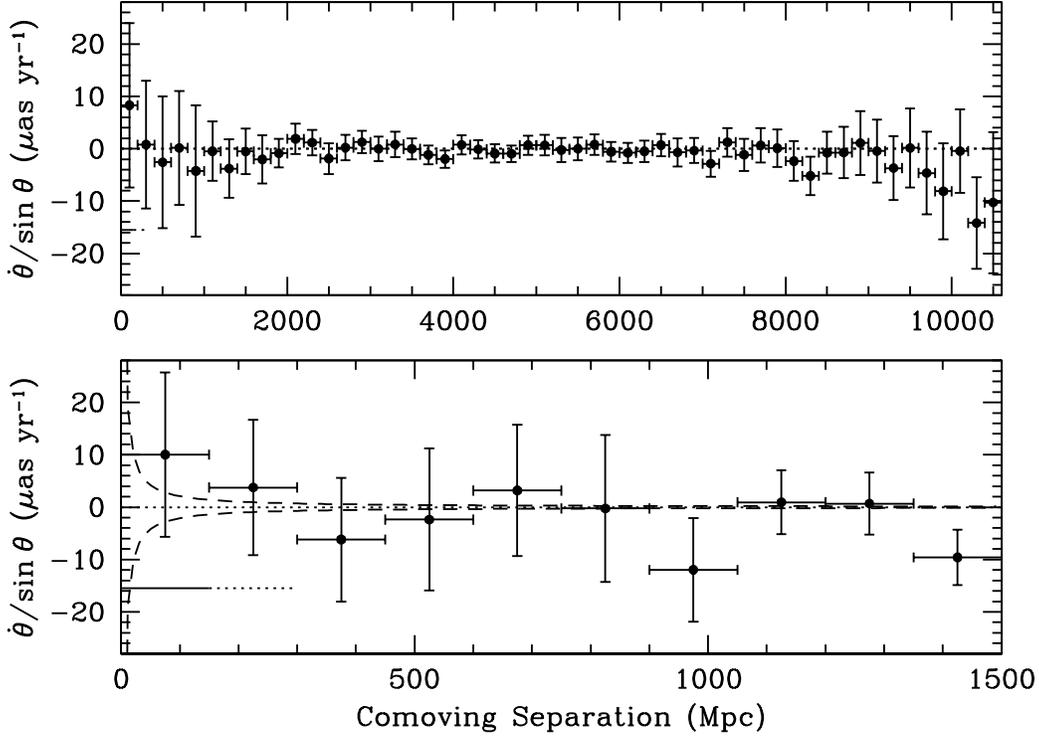}
\caption{\footnotesize
Measured pairwise divergence or convergence vs.\ comoving separation of radio sources in 
the \citet{titov11b} proper motion sample.
Top:  pairwise fractional proper motion of large-separation pairs showing no 
systematic or significant offset from pure Hubble expansion.
Bottom:  pairwise fractional proper motion of pairs with comoving separations $<1500$~Mpc.  
The line at $\dot{\theta}/\sin\theta = -15.5$~$\mu$as~yr$^{-1}$
indicates the expected convergence at $z=0$ for static structures (Equation (\ref{eqn:pm_exact}), $\dot{\ell}=0$).
The dashed lines show the effect of a peculiar transverse motion of $\pm1000$~km~s$^{-1}$ on otherwise comoving pairs.
The point with comoving separation less than 150 Mpc is consistent with Hubble expansion and 
static structures (and is therefore inconclusive given the present sample size
and measurement precision).  
}\label{fig:obs-thetadot_vs_D}
\end{figure*}

\section{Individual Structures:  Geometric Distance and the Hubble Constant}
 
Individual galaxies or galaxy clusters will have a roughly fixed physical 
size over any human-timescale observation, so $\dot{\ell} \simeq 0$.
For the local universe, $z\simeq0$, $H(z)\simeq H_\circ$, and 
\begin{equation}
 \dot{\theta} \simeq -H_\circ \theta \simeq -\theta (^\circ)\ 0.27\ \mu\rm{as\ yr}^{-1}
\label{eqn:theta-dot-static-low-z}
\end{equation}
for small angles $\theta$.  
A large nearby cluster moving with the Hubble expansion, 
such as the Perseus Cluster (Abell 426), spanning $\sim$14$^\circ$ \citep{hamden10},
would thus appear to shrink by $\sim$4 $\mu$as~yr$^{-1}$ (for comparison, the equivalent radial contraction velocity 
in a static universe would be $\sim 1400$~km~s$^{-1}$).
The Andromeda galaxy's $\sim$2$^\circ$ molecular ring, approaching at effectively 5.3 $H_\circ$ ($-$300 km s$^{-1}$ heliocentric at
780 kpc), will appear to grow by $\sim$3~$\mu$as~yr$^{-1}$ 
\citep[equivalent to a radial expansion of $\sim10$~km~s$^{-1}$;][]{darling11}.   While galaxies within clusters and individual 
maser-emitting regions within galaxies may exhibit peculiar velocities, a virialized cluster or a rotating disk galaxy will 
not exhibit a radial change in size that could be confused with the cosmological recession 
(see also Figure \ref{fig:thetadot}).  A large peculiar motion such as the initial infall expected for the Bullet Cluster, however, 
with $v_\perp=3000$~km~s$^{-1}$ and $\ell=5$~Mpc at $z=0.296$ \citep{mastropietro08}, 
would produce proper motion of $-$0.5 $\mu$as~yr$^{-1}$ compared to the 
cosmic recession of $-$0.08~$\mu$as~yr$^{-1}$.  The peculiar motion would dominate the proper motion 
in this case because the pre-Bullet Cluster had a small size, large peculiar motion, and thus large $\dot{\ell}/\ell$.

The apparent shrinking
of receding objects provides a direct geometric measurement of the Hubble constant $H_\circ$ (modulo peculiar velocity),
\begin{equation}
  H_\circ \simeq - {\,\dot{\theta}\,\over \theta}\ ,
\end{equation}
and the geometric (proper) distance, 
\begin{equation}
  D \simeq - v\, {\theta\over \dot{\theta}}\ ,
\end{equation}
which is peculiar velocity-independent
(the exact relationship is modulated by the ratio between proper distance (Hubble's Law) and angular size distance, $1+z$, but
these are very similar at $z\simeq0$).
$\theta$, $\dot{\theta}$, and Doppler velocity $v$ are observable quantities:  this 
implies that the Hubble constant and distance can be directly measured from the apparent proper motion of 
receding objects.  
These measurements do not rely on any information about the physical size or orientation of the observed shrinking object, 
in contrast to the canonical cosmological ``standard ruler'' tests. 
Moreover, since the Doppler velocity and the angular size can be measured extremely precisely, the uncertainty in 
these measurements is dominated by the uncertainty in $\dot{\theta}$.  And while the measurement
of $H_\circ$ relies on an assumption of motion entrained in the Hubble flow (small peculiar velocity), the measurement
of geometric distance does not rely on any assumptions because receding objects appear to shrink regardless of the 
reason for the recession (peculiar or cosmological).

Measuring apparent proper motions requires compact luminous (high brightness temperature) sources at the 
boundaries of the receding object or structure.  Typically these sources will be masers, which are severely distance-limited, 
or active galactic nuclei (AGNs).  In the case of galaxy clusters, bright AGNs on the periphery of clusters are 
rare; they typically reside in cluster centers.  Since quasars and galaxy clusters mark density peaks in 
large scale structure, it stands to reason that gravitationally bound or Hubble flow-decoupled structures as 
revealed by quasars and clusters of galaxies will show a relative proper motion as large scale structure decouples
from the universal expansion.

\section{Apparent Proper Motion of Large Scale Structures}
 
 Separating $\dot{\ell}/\ell$ into expansion and peculiar velocity parameterized by $\xi$, which can
be related to density contrast $\delta\rho/\langle\rho\rangle$ in the linear regime
but is otherwise a free parameter for peculiar velocity scaled to the local Hubble expansion,
\begin{equation}
   {\dot{\ell}\over\ell} = {H(z)\over1+z}\ (1-\xi).
\end{equation}
Equation (\ref{eqn:pm}) becomes 
\begin{equation}
{\dot{\theta}\over\theta} = {- H(z)\over1+z}\ \xi.
\label{eqn:pm2}
\end{equation}
Thus, if $\xi=0$ (i.e., $\delta\rho=0$), the structure expands with the Hubble flow and there is no apparent proper motion.  
If $\xi=1$, $\dot{\ell} = 0$, and $\ell$ is a cosmic ruler.
If $\xi>0$, there is apparent convergence, and 
if $\xi<0$, there is apparent divergence (i.e., voids).  For quasar pairs, $\delta\rho/\langle\rho\rangle$ is usually in 
the linear regime (typical quasar pairs are close to or much farther apart than the homogeneity scale).
Objects at the edges of voids are expected to move apart due to collapse away from voids, which 
are otherwise expanding with the Hubble flow, so a slight divergence could be expected.  This would be 
a few $\mu$as at most at low redshift and $< 1$~$\mu$as~yr$^{-1}$ at $z\gtrsim0.1$.

Figure \ref{fig:thetadot} shows the ``observer's plot'' of the expected proper motion of structures of various size scales  
and peculiar plane-of-sky velocities versus redshift along with measurements of individual quasar
pairs (Section \ref{sec:data}).  Peculiar motions are a small contribution to $\dot{\theta}$ except for 
small or nearby structures, although for large structures the quantity of relevance is the velocity gradient. 
For example, the Great Wall's $-$15~$\mu$as~yr$^{-1}$ recession-equivalent contraction velocity 
is $\sim$9000 km~s$^{-1}$, which is a velocity gradient of only about 37 km~s$^{-1}$~Mpc$^{-1}$.  In any case, 
structures, with the exception of voids, do not generally experience peculiar expansion, so gravitational contraction
enhances the proper motion signal, and the dominant consideration becomes the impact of $\xi$ on apparent contraction.  

Figure \ref{fig:thetadot-theory} shows the ``theorist's plot'' of the expected 
fractional proper motion for various $\xi$ values and 
peculiar plane-of-sky velocities versus redshift along with measurements of individual and binned quasar
pairs (Section \ref{sec:data}).  $H(z)/(1+z)$ is a slowly varying function of redshift and can be approximated
as a constant, $\sim15$~$\mu$as~yr$^{-1}$.    This plot properly shows the enhanced
effect of $v_\perp$ on smaller structures and demonstrates that small-angular-separation quasar pairs with 
precisely measured $\dot{\theta}$ are needed to make the first measurement of the cosmic recession effect.

\begin{figure*}
\epsscale{0.9}
\plotone{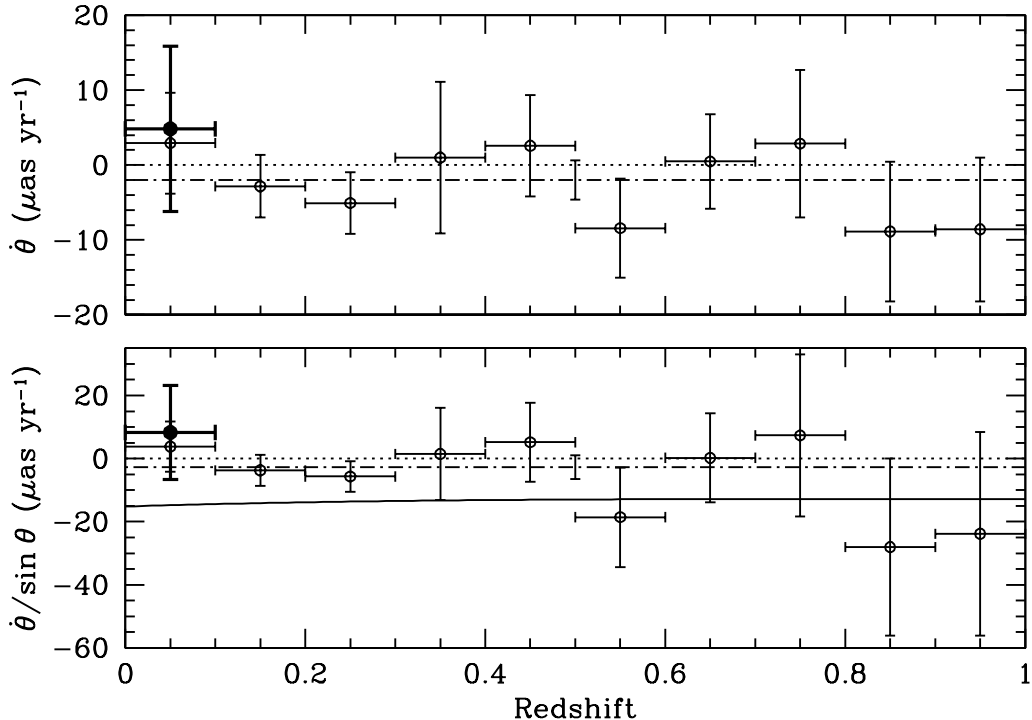}
\caption{\footnotesize
Measured pairwise divergence or convergence vs.\ redshift of radio sources in 
the \citet{titov11b} proper motion sample.
Bold filled circles indicate pairs with comoving separation less than 200 Mpc, and 
empty circles indicate pairs with comoving separation between 200 Mpc and 1500 Mpc (Figure \ref{fig:thetadot-theory}
shows all pairs and redshifts).
Top:  pairwise proper motion.
Bottom:  pairwise fractional proper motion.  The solid line 
indicates the expected convergence for static structures, $\dot{\theta}/\sin\theta = -H(z)/(1+z)$ (Equation (\ref{eqn:pm_exact}), 
$\dot{\ell}=0$).
Dot-dash lines show the mean of all large comoving separation pairs (200--1500 Mpc) in $0<z<1$,
and the small error bars at $z=0.5$ show the uncertainty in the mean, demonstrating that the large-separation 
pairs are consistent with pure Hubble expansion and inconsistent with static structures with $3.5\,\sigma$ confidence.  
The single point with 
comoving separation less than 200 Mpc, however, is consistent with Hubble expansion and with 
the expectation for static structures.
}\label{fig:obs-thetadot_vs_z}
\end{figure*}

\section{A First Application of Data}\label{sec:data}

We employ the \citet{titov11b} proper motion measurements of 555 radio sources, using the 507 with
redshifts \citep[updated online catalog]{titov11a}, to attempt a first test of the expected real-time convergence 
of Hubble flow-decoupled pairs and to confirm the pure Hubble expansion of large-separation pairs.
The data were obtained from 5030 sessions of the permanent geodetic and astrometric very long baseline interferometry (VLBI) 
program, which includes the Very Long Baseline Array,\footnote{The National Radio Astronomy Observatory is a 
facility of the National Science Foundation operated under cooperative agreement by Associated Universities, Inc.}
at 8.4 GHz in 1990--2010 using a relaxed per-session no-net rotation
constraint \citep{titov11b}.

While pairwise proper motions are minimally affected by the 
secular aberration drift caused by the barycenter acceleration about the Galactic center, we nonetheless 
subtract the dipole proper motion pattern first identified by \citet{titov11b} and confirmed by 
\citet{xu12} but employing the \citet{reid13} results, $R_0=8.38\pm0.18$~kpc and $\Theta_0=243\pm7$~km~s$^{-1}$,
which give a dipole amplitude of $5.0\pm0.3$~$\mu$as~yr$^{-1}$, from the observed proper motion vector field.
We assume that the acceleration direction is exactly toward the Galactic center and 
do not include the out-of-the-disk acceleration described by \citet{xu12} in our correction.

In order to omit poorly-measured proper motions and objects with large intrinsic proper motions,
objects used in this analysis are restricted to proper motion and uncertainty
$<100$~$\mu$as~yr$^{-1}$.  These criteria reduce the sample to 284 objects. 
Pairs of objects are likewise restricted to have $|\dot{\theta}|$ and $\sigma_{\dot{\theta}} <100$ $\mu$as~yr$^{-1}$.
While these choices are somewhat arbitrary, 
different cutoff values of the same order of magnitude have minor impact on the results.

The individual pairs with comoving separation $<150$~Mpc and small $\dot{\theta}$ in Figure \ref{fig:thetadot} are
ICRF J110427.3+381231 ($z=0.03$) and J123049.4+122328 ($z=0.004$), 
J110427.3+381231 and J163231.9+823216 ($z=0.025$), 
J110427.3+381231 and J165352.2+394536 ($z=0.034$), 
J165352.2+394536 and  J180650.6+694928 ($z=0.051$),
J123049.4+122328 and J163231.9+823216, and 
J123049.4+122328 and J165352.2+394536.  The latter two pairs also have small $\dot{\theta}/\sin\theta$ 
(Figure \ref{fig:thetadot-theory}).

Figure \ref{fig:obs-thetadot_vs_D} shows the divergence/convergence of radio source pairs (Equations (\ref{eqn:thetadot_pm}) and 
(\ref{eqn:thetadot_detail}))
versus their comoving separation.  The comoving separation is calculated from comoving proper distances using the cosine rule
and Equation (\ref{eqn:theta}):
\begin{equation} 
r_{ij}^2 = r_i^2+r_j^2-2 r_i r_j \cos\theta_{ij}.
\end{equation}
Error bars are estimated from bootstrap resampling.  
The expected signal for static structures ($\dot{\ell}=0$) at $z=0$ is $\dot{\theta}/\sin\theta = -15.5$~$\mu$as~yr$^{-1}$
(Equation (\ref{eqn:pm_exact})).
The data point with comoving separation $<150$~Mpc is consistent with this value 
($\dot{\theta}/\sin\theta=+10\pm16$~$\mu$as~yr$^{-1}$; 1.6$\,\sigma$ separation), but it is also consistent with 
pure Hubble expansion ($\dot{\theta}=0$); more pairs or precision are needed.  
Pairs with comoving separations 150--1500 Mpc are consistent with pure Hubble expansion, as expected, as are those with
separations 0--1500 Mpc, which are inconsistent with the signal from static structures at
$z=0$ with 3.7$\,\sigma$ significance: $\langle\dot{\theta}/\sin\theta\rangle = -2.3\pm3.6$~$\mu$as~yr$^{-1}$ and
$\langle\dot{\theta}\rangle = -1.7\pm2.4$~$\mu$as~yr$^{-1}$ (Table \ref{tab:bins}).

Figure \ref{fig:obs-thetadot_vs_z} shows the divergence/convergence of pairs 
versus their mean redshift, grouped into two populations:  those with comoving separations less than 200 Mpc, and those
with comoving separations between 200 and 1500 Mpc (Table \ref{tab:bins}; Figure \ref{fig:thetadot-theory} shows all pairs).
Because the bright radio sources suitable for proper motion measurements have a low areal density, 
the small-separation pairs are necessarily at low redshift ($z<0.1$) where large angles can span small proper distances.
For large-separation pairs, the redshift difference between pairs can be larger than the $\Delta z=0.1$ redshift bin, and 
binned redshifts are averages, not the redshifts corresponding to averaged distances.
Static structures should show $\dot{\theta}/\sin\theta = -15.5$~$\mu$as~yr$^{-1}$ 
at $z=0$, and this evolves slowly with redshift.  
The sole data point with small comoving separation ($<200$ Mpc) is consistent with this value,  
$\dot{\theta}/\sin\theta=+8.3\pm14.9$~$\mu$as~yr$^{-1}$ or 1.6$\,\sigma$ deviation, but it is also consistent with 
pure Hubble expansion, $\dot{\theta}/\sin\theta=0$.  More small-separation pairs or precision are needed.
Pairs with comoving separations 200--1500 Mpc at $z<1$ are consistent with pure Hubble expansion, as expected:
$\langle\dot{\theta}/\sin\theta\rangle = -2.7\pm3.7$~$\mu$as~yr$^{-1}$ and
$\langle\dot{\theta}\rangle = -2.0\pm2.6$~$\mu$as~yr$^{-1}$.
This is inconsistent with the signal from static structures at $z=0$ at 3.5$\,\sigma$ significance.

\begin{deluxetable}{ccrr}
\tabletypesize{\scriptsize} 
\tablecaption{Binned Pairwise Proper Motions \label{tab:bins}} 
\tablewidth{0pt} \tablehead{
\colhead{Comoving} &
\colhead{$\langle z \rangle$} & 
\colhead{$\langle\dot{\theta}\rangle$} &
\colhead{$\langle\dot{\theta}/\sin\theta\rangle$} \\
\colhead{Separation} &
\colhead{} & 
\colhead{} & 
\colhead{} \\
\colhead{(Mpc)} & \colhead{} & \colhead{($\mu$as yr$^{-1}$)} & \colhead{($\mu$as yr$^{-1}$)}}
\startdata 
{\bf 0--200}   & {\bf 0.0--0.1} & {\bf 4.8(11.0)} & {\bf 8.3(14.9) }\\
200--1500  & 0.0--0.1 &       2.9(6.7) &       3.8(7.9) \\
200--1500  & 0.1--0.2 & $-$2.8(4.2) & $-$3.7(5.0) \\
200--1500  & 0.2--0.3 & $-$5.1(4.1) & $-$5.6(4.8) \\
200--1500  & 0.3--0.4 &     1.0(10.1) &       1.5(14.7) \\
200--1500  & 0.4--0.5 &       2.6(6.8) &       5.2(12.5) \\
200--1500  & 0.5--0.6 & $-$8.4(6.6) &$-$18.6(15.8) \\
200--1500  & 0.6--0.7 &       0.5(6.3) &        0.2(14.1) \\
200--1500  & 0.7--0.8 &       2.9(9.8) &        7.4(25.7) \\
200--1500  & 0.8--0.9 & $-$8.9(9.3) &$-$28.1(28.1) \\
200--1500  & 0.9--1.0 & $-$8.6(9.6) &$-$23.8(32.3) \\
\noalign{\vskip 1mm}
\hline
\noalign{\vskip 1mm}
 200--1500  & 0.0--1.0 & $-$2.0(2.6) & $-$2.7(3.7) \\
\noalign{\vskip 1mm}
\hline
\noalign{\vskip 1mm}
{\bf 0--150} & {\bf Any} &  {\bf  6.7(8.9)} & {\bf 10.0(15.7)} \\
150--300 & Any &       2.1(11.2) &     3.8(12.9) \\
300--450 & Any & $-$3.3(7.4) & $-$6.2(11.8)\\
450--600 & Any & $-$5.0(8.2) & $-$2.4(13.6)\\
600--750 & Any &       0.4(6.8) &       3.2(12.6)\\
750--900 & Any &       2.7(7.5) & $-$0.2(14.0)\\
900--1050 & Any &$-$6.9(5.6)& $-$12.0(9.9)\\
1050--1200 & Any &    0.9(4.5)&         0.9(6.1)\\
1200--1350 & Any &$-$0.0(4.5)&       0.7(5.9)\\
1350--1500 & Any &$-$6.1(3.5)& $-$9.6(5.3) \\
\noalign{\vskip 1mm}
\hline
\noalign{\vskip 1mm}
 0--1500 & Any & $-$1.7(2.4) & $-$2.3(3.6)\\
\noalign{\vskip 1mm}
\hline
\noalign{\vskip 1mm}
 Any & Any & $-$0.35(0.52) & $-$0.36(0.62)
\enddata 
\tablecomments{Bold entries indicate comoving
separations where proper motions are expected to 
deviate from pure Hubble flow.  Parenthetical values are $1\,\sigma$ uncertainties.
}
\end{deluxetable}

Using all pairs and all redshifts (Figure \ref{fig:thetadot-theory}), we reject the static locus at 
$\sim$5--10$\,\sigma$ significance for $z\simeq0.5$--2.  
 Likewise, for the entire sample unbinned in redshift
  we obtain a ``null test'' pairwise proper motion of $\langle\dot{\theta}/\sin\theta\rangle =-0.36\pm0.62$~$\mu$as~yr$^{-1}$
  and $\langle\dot{\theta}\rangle =-0.35\pm0.52$~$\mu$as~yr$^{-1}$,
  consistent with pure Hubble expansion (the negligible difference between $\dot{\theta}$ and 
$\dot{\theta}/\sin\theta$  is due to the 
 angular separation averaging to 90$^\circ$; the two values no longer match if one does not subtract the aberration drift signature
 from the proper motion vector field).
  These precise measurements are possible despite the large apparent proper motions intrinsic to radio jets because intrinsic 
 motions are not correlated  between objects. 

\vspace{30pt}

\section{Discussion}

As Figures \ref{fig:obs-thetadot_vs_D} and \ref{fig:obs-thetadot_vs_z} show, the expected pairwise convergence 
effect should be detectable using current angular resolution, astrometry, and proper motion sensitivity.  
The major impediment to progress is the limited number of close quasar pairs.  The binned large-separation pairs can 
reach uncertainties of $\sim1$~$\mu$as~yr$^{-1}$, which is more than adequate to detect convergence and recession of 
structures were similar numbers of sub-150 Mpc pairs observed.  
Higher bandwidth VLBI recording 
can grow the radio proper motion sample by an order or magnitude, but the large intrinsic proper motions 
manifested in many radio sources will still be a limitation.  Optical proper motions obtained by
the {\it Gaia} mission\footnote{ http://www.rssd.esa.int/SYS/docs/ll\_transfers/project$=$PUBDB\&id$=$448635.pdf} 
will benefit from a vastly larger 
sample of $\sim$500,000 quasars and from negligible
intrinsic proper motion.  {\it Gaia} will achieve astrometry of $\sim$80 $\mu$as for $V=18$ mag stars \citep{debruijne05}.

Future observations, whether radio or optical, should be able to detect the statistical convergence signal and 
may detect the recession effect in single nearby pairs as well.  Individual low-redshift pairs in the \citep{titov11b}
are sample already within a factor of a few of the precision needed to test the recession effect (Figure \ref{fig:thetadot-theory}), 
and the geodetic observations were not designed for this purpose.  
A true ``moving cluster'' observation of a galaxy cluster
may someday be possible, providing a geometric distance from cosmic expansion alone.

\section{Conclusions}

While the sample of small-separation quasar pairs with precise proper motion measurements is as-yet too sparse
to detect the cosmic recession and collapse of structure, 
large-separation test masses have now been measured with high significance to
be comoving with the Hubble expansion
and can serve as a reference frame for small-separation pairs that will significantly deviate
from the Hubble flow due to gravity.  This relative measurement of small-separation versus large-separation 
quasar pairs will mitigate possible systematic effects inherent in such precise proper motion measurements 
given the large intrinsic proper motions seen in radio sources.  Improved VLBI astrometry and the {\it Gaia}
astrometry mission will likely detect the departure of structures from pure Hubble expansion in a statistical
sample as well as for individual structures.  It may also be possible to obtain geometric distances and measure the
Hubble constant by observing relatively static objects such as individual galaxies or galaxy clusters.  

\acknowledgments
The author thanks \citet{wright06} for the online cosmology calculator and the anonymous referee for 
helpful comments.
This research has made use of the NASA/IPAC Extragalactic Database (NED) which is operated 
by the Jet Propulsion Laboratory, California Institute of Technology, under contract with NASA.

\end{document}